\documentclass[10pt]{article}
\usepackage{conf_iap}

\begin{document}
\begin{flushright}
CERN-TH/2002-246 
\end{flushright} 
\setcounter{footnote}{0}
\heading{Accelerated expansion without dark energy}
\par\medskip\noindent
\setcounter{footnote}{0}
\author{Dominik J. Schwarz$^{1}$}
\address{Theory Division, CERN, 1211 Geneva 23, Switzerland \\ 
dominik.schwarz@cern.ch} 

\begin{abstract}
The fact that the $\Lambda$CDM model fits the observations 
does not necessarily imply the physical existence of `dark energy'.
Dropping the assumption that cold dark matter (CDM) is a perfect fluid
opens the possibility to fit the data without dark energy. For imperfect 
CDM, negative bulk pressure is favoured by thermodynamical arguments
and might drive the cosmic acceleration. The coincidence between 
the onset of accelerated expansion and the epoch of structure formation 
at large scales might suggest that the two phenomena are linked.
A specific example is considered in which effective (anti-frictional)
forces, which may be due to dissipative processes during the formation of 
inhomogeneities, give rise to accelerated expansion of a CDM universe. 
\end{abstract}
\bigskip 

\noindent {\it Talk given at the XVIIIth IAP Colloquium
``ON THE NATURE OF DARK ENERGY: Observational and theoretical results on 
the accelerating universe'', Institut d'Astrophysique de Paris, France, 
July 1 -- 5, 2002.} 
\bigskip

\section{How many dark components has the Universe?}

It might appear that overwhelming evidence for the existence of dark energy 
(be it a cosmological constant or some other form of energy) has been collected.
Observations of the cosmic microwave background (CMB) \cite{CMB}, high redshift 
supernovae \cite{SN}, the large scale structure \cite{LSS} and clusters 
\cite{C} give a self-consistent picture, which can be summarized 
by the dimensionless energy densities $\Omega \approx 1, 
\Omega_\Lambda \approx 0.7, \Omega_{\rm b} h^2 \approx 0.02, 
\Omega_{\rm cdm} h^2 \approx 0.12$, 
implying $h \approx 0.7$ and an age of the Universe of $14$ Gyr. But this 
evidence for dark energy is obtained from a fit to 
the $\Lambda$CDM model, not by a direct observation of dark energy itself. 

In this work we point out that the $\Lambda$CDM model makes 
a very important, yet untested assumption: CDM\footnote{
Any dark matter with $|P| \ll \rho$ during the epoch of structure
formation is called CDM.} is described by a perfect 
fluid (at the largest scales). We show below that if this 
assumption were wrong, the evolution of the Hubble rate would have to 
be modified. 

Let us first check whether there are any theoretical or observational 
reasons for believing that CDM behaves like a perfect fluid.
The observed isotropy of the CMB together with the 
Copernican principle gives sound evidence that the Universe is 
homogeneous and isotropic at very large scales (cosmological principle). 
This implies that the energy-stress tensor of the Universe is of the form
\begin{equation}
\label{emt}
T^a_b = (\rho + P)u^a u_b + P \delta^a_b,
\end{equation}
where $u^a$ is the velocity of an observer comoving with the CMB heat bath,
$\rho$ and $P$ are the energy density and hydrodynamic pressure 
as measured by this observer. Note that this form does not imply that the 
fluid is perfect, i.e.~dissipationless. However, the observed CMB spectrum 
proves that the radiation component is in thermal equilibrium. Assume for
the moment that CDM consists of weakly interacting massive particles (WIMPs)
that were in thermal equilibrium with the radiation in the early Universe.
For the most popular WIMP, the neutralino, kinetic decoupling from the 
radiation fluid happens at a temperature of about $10$ MeV \cite{HSS},
long before structure formation starts. After kinetic decoupling 
any perturbation can easily disturb the CDM equilibrium distribution.  
A general result from kinetic theory within general relativity is that it is
very hard to maintain kinetic equilibrium for freely streaming particles.
This is possible only if a conformal time-like Killing vector exists and the 
particles are either massless or highly non-relativistic \cite{Ehlers}. 
For CDM these conditions are approximately true until density  
perturbations become non-linear, and it is therefore well justified (at least 
for WIMPs that once were in thermal contact) to use a model with 
radiation plus CDM at the beginning of structure formation and during 
photon decoupling. 

But today's Universe is very inhomogeneous on smaller scales, and we certainly 
cannot describe CDM by a perfect fluid on those scales. All kinds of 
dissipative effects might take place. Now the question arises whether looking 
at larger scales only (this is nothing but averaging over the smaller scales)
can justify the assumption of a perfect fluid. A simple argument shows why 
this cannot be true in general. Assume that entropy is created by the 
dissipative effects in every small volume. Averaging over many small volumes 
will only add up the produced entropy, the net result being that entropy is 
produced at large scales as well, although no physical processes act at  
the large scales themselves.
 
To take into account dissipative effects in the formation of small scale 
inhomogeneities on the cosmic evolution on larger scales, one can start 
from the energy-stress tensor of an imperfect fluid (\ref{emt}). 
If dissipative processes are taking place the hydrodynamic pressure $P$ 
no longer equals the kinetic pressure $p$, and we write $P = p + \Pi$. 
For CDM $p \approx 0$. We argue below that, if $\Pi \neq 0$, 
the observational evidence for two different dark components of the Universe 
breaks down. This is consistent with Pav\'on's contribution to this conference,
in which he shows that in order to have late-time cosmic acceleration and to 
solve the coincidence problem at the same time, matter must be dissipative 
\cite{Diego}, although a quintessence component is admitted in that work.  

What can be said about the sign of $\Pi$? Let us assume for simplicity that 
the energy density of baryons is much less than that of dark matter, so all 
dynamically important energy density is in CDM.
As long as the baryon number is conserved, we can define the entropy per baryon
(specific entropy) $\sigma$, and the relation
\begin{equation}
T {\rm d}\sigma = {\rm d}(\rho/n_{\rm B}) + p {\rm d}(1/n_{\rm B})
\end{equation}
holds true for any change of the thermodynamic state. Note that it is 
the kinetic pressure that enters in this expression. Using the
covariant conservation of energy density and the baryon number conservation,
we find the change of the heat density with time 
\begin{equation}
n_{\rm B} T \dot{\sigma} = - 3 H \Pi,
\end{equation}
where $H$ denotes the Hubble rate. From the second law of thermodynamics
$\Pi \leq 0$ follows for an expanding Universe, and in the case 
of CDM ($p \approx 0$) we find $P \approx \Pi \leq 0$! Thus a negative CDM bulk 
pressure seems possible from non-equilibrium effects during structure 
formation. A classic example in which the non-equilibrium pressure
reduces the kinetic pressure is the bulk viscosity of cosmic fluids 
in the linear approximation, $\Pi = - 3 H \zeta < 0$ \cite{Weinberg,PZ}.  

Let us now consider the hypothesis that the Universe contains just 
one CDM component, which is described as an imperfect fluid at late times
(redshifts of a few). Such a solution to the dark energy problem would
be most elegant because no new particles or fields should be introduced and
no new physics, such as extra dimensions or new forces, are needed.
Under this hypothesis the cosmic coincidence problem turns into the question: 
Why do cosmic acceleration and the formation of large structures happen at 
about the same time? If the non-linear evolution of inhomogeneities could
be identified as the driver for cosmic acceleration, the coincidence 
would be explained naturally \cite{DJS}.

\section{A CDM model with antifriction}

In order to be able to describe the cosmological observations, we need 
a model for the evolution of $\Pi$. Let us try to obtain such a
model from a microscopic ansatz. In every-day physics dissipative 
effects can be described by effective forces such as the Stokes friction. 
Indeed, the only force that respects the cosmological principle
is an effective (anti-)frictional force, which reads in the Newtonian limit 
\begin{equation}
\vec{F} \approx  - B(m,t) m \vec{v},
\end{equation}
where $m$ is the mass of a test particle and $\vec{v}$ its peculiar velocity 
\cite{ZSBP}; $B$ is the coefficient of friction, which has the dimension 
of a rate. The kinetic theory incorporating such an effective friction term 
has been discussed in Ref.~\cite{ZSBP}. Under the assumption
that the distribution function is in a generalized equilibrium, the dynamic  
pressure can be calculated to be
\begin{equation}
P \approx (B/H)\rho .
\end{equation}
Thus, for effective antifrictional forces ($B<0$) the dynamic pressure of CDM 
is negative. Note that the number of test particles is not conserved in 
generalized equilibrium, see \cite{ZSBP}. 

Guided by the idea that CDM becomes an imperfect fluid as structures 
in the Universe grow non-linear, one can estimate the rate of antifriction
from a dimensional argument. The only rate in the problem is the 
Hubble rate, thus $B(t_0) = - \nu H_0$, $\nu$ being a constant of order unity.
For the time dependence we have tested three different ans\"atze in 
Ref.~\cite{ZSBP}. 
It turns out that all three of them can provide excellent fits to the 
supernova data, all with $\nu ={\cal O}(1)$. This is already surprising, 
but what is even more surprising is that the ansatz 
$B(z) = - \nu (H_0/H) H_0$
gives rise to a model that is dynamically equivalent to the $\Lambda$CDM model.
This shows that CDM with antifriction fits the supernovae, CMB and  
large scale structure observations as good as the $\Lambda$CDM model itself. 

\section{Conclusions}

It seems that a careful investigation should be done to sort out 
the number of dark components of the Universe. We have shown that 
a fit to the $\Lambda$CDM model cannot be sufficient evidence, as 
long as the assumption that CDM is a perfect fluid at large scales 
remains untested.

A possible argument against the proposal of this paper might be 
that observations of galaxy clusters indicate $\Omega_{\rm m} \sim 0.3 < 1$!
The discrepancy can be resolved by taking into account that the gravitating 
mass density is $\rho + 3P$, and therefore cluster mass estimates
probe $\Omega_{\rm m}(1 + 3P/\rho)$ rather than $\Omega_{\rm m}$. Only 
geometrical mass estimates probe $\Omega$ directly (as in the case of the CMB).
On the largest scales (CMB, large scale structures and supernovae) 
$\Omega_{\rm m} =1$ would imply that $P \sim  - 0.7 \rho$, whereas on 
cluster scales the observed low mass density would be consistent with 
$P \sim - 0.2 \rho$. Thus an observational signature of the present scenario 
is a time- and scale-dependent effective equation of state of CDM \cite{DJS}.  
These and other questions should be addressed in detail, before 
the energy budget of the Universe is understood. 

\acknowledgements{I wish to thank the organizers for their support, and 
it is a pleasure to acknowledge discussions and collaboration with 
A.~Balakin, D.~Pav\'on and W.~Zimdahl.}

\vfill
\end{document}